\documentclass[aps,onecolumn,showpacs,floatfix,nofootinbib,superscriptaddress,11pt]{revtex4-1}
\newcommand{\beq}{\begin{eqnarray}}
\newcommand{\eeq}{\end{eqnarray}}
\newcommand\<{\langle}
\renewcommand\>{\rangle}
\renewcommand\d{\partial}

\renewcommand\d{\partial}

\usepackage{amssymb,amsmath}
\usepackage{graphicx,float,color}
\usepackage{indentfirst}
\newcommand{\be}{\begin{equation}}
\newcommand{\ee}{\end{equation}}

\newcommand{\ba}{\begin{eqnarray}}
\newcommand{\ea}{\end{eqnarray}}

\newcommand{\GN}{G_{\rm N}}

\newcommand\op[1]{{\cal A}_{#1}^{\mu\nu}}
\begin{document}
\title{Fluid/gravity correspondence and the CFM black brane solutions}
\author{R.~Casadio}
\email{casadio@bo.infn.it}
\affiliation{Dipartimento di Fisica e Astronomia, Universit\`a di Bologna
\\
via Irnerio~46, I-40126 Bologna, Italy}
\affiliation{I.N.F.N., Sezione di Bologna, \\ via B.~Pichat~6/2, I-40127 Bologna, Italy}
\author{R.~T.~Cavalcanti}
\email{rogerio.cavalcanti@ufabc.edu.br}
\affiliation{Dipartimento di Fisica e Astronomia, Universit\`a di Bologna
\\
via Irnerio~46, I-40126 Bologna, Italy}
 \affiliation{
Centro de Ci\^encias Naturais e Humanas, 
Universidade Federal do ABC - UFABC,\\ 09210-580, Santo Andr\'e, Brazil}

\author{Rold\~ao~da~Rocha}
\email{roldao.rocha@ufabc.edu.br}
\affiliation{Centro de Matem\'atica, Computa\c c\~ao e Cogni\c c\~ao, 
Universidade Federal do ABC - UFABC,\\ 09210-580, Santo Andr\'e, Brazil.}
\pacs{11.25.Tq, 11.25.-w, 04.50.Gh} 
\begin{abstract} 
We consider the lower bound for the shear~viscosity-to-entropy density ratio, 
obtained from the fluid/gravity correspondence, in order to constrain the post-Newtonian
parameter of brane-world metrics.
{
In particular, we analyse the Casadio-Fabbri-Mazzacurati (CFM) effective solutions
for the gravity side of the correspondence and argue that  including higher order
terms in the hydrodynamic expansion can lead to a full agreement with the experimental
bounds, for the Eddington-Robertson-Schiff post-Newtonian parameter in the CFM metrics}. 
This lends further support to the physical relevance of the viscosity-to-entropy ratio lower
bound and fluid/gravity correspondence. Hence we show that CFM black branes are,  
effectively, Schwarzschild black branes.
\end{abstract}\maketitle
%
%
%
\section{Introduction}
Black hole solutions of the Einstein equations in general relativity (GR) are useful tools
to investigate the space-time structure and quantum effects in any possible theory of gravity.
In particular, models with extra spatial dimensions~ are prominent candidates
as extensions of Einstein GR \cite{maartens}, leading to important consequences, not only for black hole
physics~\cite{Antoniadis:1998ig,Antoniadis:1990ew,landsberg}, but also for particle physics, cosmology, as well as for the astrophysics
of supermassive objects~\cite{kuerten2014}.
In this context, black strings play a notable role in the quest for physically viable and realistic
models~\cite{clark,cor2014,rr22014}.
\par
Fluid dynamics is an effective description of interacting quantum field theories (QFTs) at long
wavelengths~\cite{Bu:2014ena}.
As a low energy effective field theory, fluid dynamics is determined by a derivative expansion
of the local fluid variables, and is used to describe near equilibrium systems.
The derivative expansion is determined by the symmetries of the system and its 
thermodynamical features.
Transport coefficients, such as viscosities and conductivities, assess how perturbations propagate
under relaxation and, besides being theoretically predictable, they can be measured experimentally.
\par
The AdS/CFT correspondence, which relates the coupled dynamics of $SU(N)$ gauge theories
to gravity in anti-de Sitter (AdS) spaces~\cite{Maldacena:1997re,Gubser:1998bc,Aharony:1999ti,Natsuume:2014sfa},
has been used to connect those two realms of modern physics, namely gravity and QFT.
It holographically relates  a strongly interacting QFT at large $N$ with the dynamics provided by  gravity
in asymptotically AdS space-time.
In the long wavelength limit, the fluid/gravity correspondence has been  recently proposed
as a framework related to the AdS/CFT correspondence~\cite{Policastro:2001yc,poli2, KSS},
mapping black holes in asymptotically AdS space-times to the fluid dynamics of a strongly coupled
boundary field theory.
One of the most remarkable predictions of AdS/CFT and fluid/gravity correspondence
is the ratio of the shear viscosity-to-entropy density, which is universal for a large class
of strongly coupled (gauge theory) isotropic plasmas~\cite{Bu:2014ena}.
The universality of this ratio plays a prominent role for gauge theories that are dual to
gravitational backgrounds~\cite{Emparan:1997iv,Policastro:2001yc,poli2,KSS,KSS0,Buchel:2004di,Cremonini:2011iq}.
Moreover, the same results were obtained by employing the Green-Kubo formula in
Ref.~\cite{Cai:2009zv}.
\par
On the other hand, black strings present hydrodynamic features such as viscosity, diffusion
rates, diffusion constants and other transport coefficients~\cite{Lehner:2010pn}, besides 
temperature and entropy.
From the perspective of the holographic principle~\cite{Maldacena:1997re}, a black string
corresponds to a certain finite-temperature QFT in lower dimensions.
In this context, the hydrodynamic features of the horizon of a black string can be identified with
the hydrodynamic behaviour of a dual theory.
For such QFTs, the ratio of the shear viscosity $\eta$ to entropy density $s$ is bounded by 
the universal value~\cite{KSS0}
\be
\frac\eta s
\gtrsim 
\frac\hbar{4\,\pi\, k_{\rm B}}
\simeq
6.08\times10^{-13}~\textrm{K} \,\textrm{s}
\ ,
\ee
for an extended range of thermal QFTs.
The above inequality holds for all known substances, including, for example, liquid helium, water, and the quark-gluon plasma, produced at the Relativistic Heavy Ion Collider (RHIC).
Therefore, it has been conjectured to represent a universal lower bound for all materials~\footnote{We shall
mostly use units with $\hbar=c=k_{\rm B}=1$, four-dimensional indices will be represented by 
Greek letters, $\mu=0,\ldots,3$ and five-dimensional indices by capital latin letters,
$A=0,\dots,4$.}, 
\be
\frac{\eta}{s}
\gtrsim 
\frac{1}{4\pi}
\ ,
\label{kss}
\ee
which is called the Kovtun-Starinets-Son (KSS)~bound~\cite{KSS,KSS0}. In the case of Einstein gravity, the ratio $\eta/s$ equals $
1/4\pi$. 
This bound is obtained in bulk supergravity from a recent calculation in ${\cal N}=4$ supersymmetric $SU(N_c)$
Yang-Mills theories, in the regime of infinite $N_c$ and large 't Hooft coupling
$g^2N_c$~\cite{Buchel:2004di}, yielding
\begin{equation}
\label{nccc}
\frac{\eta}{s}
=
\frac{1}{4\, \pi }
\left[1 + \frac {135\,\zeta(3)}{8 (2 g^2 N_c)^{3/2}} + \cdots \right]
\ ,
\end{equation}
where $\zeta (3)$ is the Ap\'{e}ry constant, and the next-to-leading term represents
the first string theory correction to GR.  
The KSS bound is a powerful tool for studying strongly interacting
systems, like the quark-gluon plasma, and is further employed to analyse trapped
atomic gases.
Strongly interacting Fermi gases of atoms have been noticed to be ruled by
hydrodynamics~\cite{OHara}, presenting finite shear viscosity at finite temperature.
The bound~\eqref{kss} holds in relativistic QFTs at finite temperature and 
chemical potential~\cite{KSS,KSS0}, for at least a single component non-relativistic gas of
particles with either spin zero or spin 1/2 ~\cite{cherman}.
The entropy of the dual QFT equals the entropy of a black string, which is
proportional to the area $A$ (volume) of its event horizon in Planck units,
$S = {A}/{4G_5}$, where $G_5$ is the five-dimensional gravitational constant. 
\par
According to the AdS/CFT duality, any given asymptotically AdS bulk geometry should be 
equivalent to specific states of the gauge theory on the boundary.
In fact, the AdS bulk geometry can be mapped into vacuum states of the gauge theory,
whereas AdS-Schwarzschild black holes are related to thermal states of the corresponding  gauge theory
on the boundary. Hence, a fluid can be associated with the bulk black hole solution.
Moreover, the bulk dynamics is governed by Einstein's equations, with the AdS-Schwarzschild
black hole as a typical solution~\cite{hubeny}.
Each different solution is equivalent to a thermal state, corresponding to  black strings
in AdS~\cite{sadeghi}.
The dynamical framework can be realised by a system in local thermodynamical equilibrium,
corresponding to a non-uniform black string in AdS, which evolves according to effective 
hydrodynamical equations corresponding to the Einstein's equations~\cite{hubeny}.
Such a system will subsequently relax towards a global equilibrium state, corresponding to a
uniform black string in AdS~\cite{sadeghi}.
\par
Motivated by the above picture, we shall here explore the fluid/gravity duality for
the five-dimensional black string~\cite{kuerten2014,Casadio:2002uv,bs1} associated with the effective
four-dimensional Casadio-Fabbri-Mazzacurati (CFM) solutions~\cite{cfabbri}. 
CFM metrics generalise the Schwarzschild solution and have a parametrised post-Newtonian
(PPN) form with parameter $\beta$, which is the usual Eddington-Robertson-Schiff parameter used to describe
the classical tests of GR~\footnote{As usual~\cite{maartens}, the four-dimensional brane-world black hole is
viewed as a transverse section of the five-dimensional black string, producing real world post-Newtonian
corrections.}.
This parameter allows for a direct comparison with experimental data in the weak-field limit,
which is accurate enough to include most solar system tests~\cite{will}. 
{In general, one can construct a dual fluid stress tensor by solving the bulk Einstein
equations in asymptotically AdS spacetime and taking the long wavelength limit, but there are only a few
black hole solutions that admit such a hydrodynamic dual on the boundary, even asymptotically.
The prototype is the Schwarzschild metric, however the CFM solutions can also be obtained as
four-dimensional vacuum brane solutions of Einstein gravity with a cosmological
constant.
In a second-order derivative theory of five-dimensional gravity interacting with other fields in AdS,
for instance type II~B supergravity in AdS$_5\times S^5$, the CFM solutions can be realised as
universal sub-sectors, regarding pure gravity with negative cosmological constant, in the
long-wavelength limit.}
\par
{On the gravity side, we shall thus consider a five-dimensional CFM black string,
associated with a four-dimensional CFM black hole, and the four-dimensional dual fluid description
will necessarily carry an imprint of the parameter $\beta$.
The fluid/gravity correspondence will then be shown to provide a bound for $\beta$, 
corresponding to the universal KSS bound~\eqref{kss}. 
In particular, both the shear viscosity and black string entropy will depend on $\beta$,
and Eq.~\eqref{kss} will read 
\be
\label{kssm}
\frac{\eta(\beta)}{s(\beta)}\gtrsim \frac{1}{4\pi}
\ ,
\ee
which is precisely what gives rise to the bound on $\beta$.}
We shall also argue that this bound can be made to agree with observations, provided higher
order terms are included in the hydrodynamic expansion of the stress tensor.
\par
This paper is organised as follows: in Section~\ref{S2} the CFM solutions are reviewed;
in Section~\ref{S3}, the Green-Kubo formula is used to bound the PPN parameter through
the shear viscosity-to-entropy density ratio in the linear regime, and we show how the estimates
improve by including higher-order hydrodynamical terms in Section~\ref{S4}; finally we discuss our results
in Section~\ref{S5}. 
\section{Casadio-Fabbri-Mazzacurati brane-world solutions}
\label{S2}
Solutions of the Einstein field equations on the brane are not uniquely determined
by the matter energy density and pressure, since gravity can propagate into the bulk
and generates a Weyl term onto the brane itself. 
Taking that into account, the CFM metrics~\cite{Casadio:2002uv} are vacuum brane solutions,
like the tidally charge metric in Ref.~\cite{Dadhich}, and contain a PPN parameter $\beta$, 
which can be measured on the brane.
The case $\beta = 1$ corresponds to the exact Schwarzschild solution on the brane, and
extends into a homogeneous black string in the bulk.
Furthermore, it was observed in Ref.~\cite{cfabbri} that $\beta \approx 1$, in solar system
measurements.
More precisely, the deflection of light in the classical tests of GR provides the bound
$|\beta-1|\lesssim 0.003$~\cite{will}.
The parameter $\beta$ also measures the difference between the inertial mass
and the gravitational mass of a test body, besides affecting the perihelion shift and describing the
Nordtvedt effect~\cite{cfabbri}. 
Finally, measuring $\beta$ provides information regarding the vacuum energy of the brane-world or,
equivalently, the cosmological constant~\cite{maartens,cfabbri}. 
\par
We recall the effective four-dimensional Einstein equations on the brane can be expressed as
\begin{eqnarray}
\label{123}
G_{\mu\nu}
=
8\,\pi\,\GN\,T_{\mu\nu}
-\frac{\Lambda_4}{2}\,g_{\mu\nu}
+ \frac{\kappa_5^4}{4}
\left[\frac{g_{\mu\nu}}{2}\left(T^2 - T_{\alpha\beta}T^{\alpha\beta}\right)+T\,T_{\mu\nu} - T_{\mu \alpha} T^\alpha _{ \;\,\nu}\right]
- {\cal E}_{\mu\nu}
\ ,
\end{eqnarray}
where $\kappa_5^2=8\,\pi\,G_5$ and $\GN=\kappa_5^2\,\sigma/48\,\pi$ ($\sigma$ being the brane tension),
$T_{\mu\nu}$ denotes the energy-momentum tensor of brane matter, $T\equiv T^{\mu}_{\;\,\mu}$,
and ${\cal E}_{\mu\nu}$ is the Weyl tensor term.
For brane vacuum, $T_{\mu\nu} = 0$, and absorbing the bulk cosmological constant $\Lambda_4$ into
the bulk warp factor, the above field equations reduce to 
\be
R_{\mu\nu}= -{\cal E}_{\mu\nu}
\ .
\label{0fe}
\ee
We are in particular interested in static spherically symmetric systems on the brane, whose general metric can be
written as
\beq
g_{\mu\nu}\,dx^{\mu}\,dx^{\nu}
=
-N(r)\,dt^2+B(r)\,dr^2 + r^2\,d\Omega^2
\ .
\label{cmetric}
\eeq
The CFM metrics are obtained by relaxing the condition $N(r)=B^{-1}(r)$, valid, for example, for the Schwarzschild and
Reissner-Nordstr\"om metrics.
In fact, for black strings, such a condition will result in a central singularity extending
all along the extra dimension and a singular bulk horizon~\cite{cfabbri}, a configuration which moreover
suffers of the well-known Gregory-Laflamme instability~\cite{gregory}.
The CFM solution~I is obtained, by fixing $N(r)$ equal to the Schwarzschild form and then determining
$B(r)$ from the field equations~\eqref{0fe}, whereas the CFM solution~II follows from the same procedure
but starting from a metric coefficient $N(r)$ of the Reissner-Nordstr\"om form~\cite{Dadhich}.
\subsection{CFM solution I}
\label{3.1}
For the first case, the metric coefficients in Eq.~\eqref{cmetric} are given by
\beq
N_I(r)
=
1- \frac{2\,\GN\, M}{r}
\qquad
\text{and}
\qquad
B_I(r)
=
\frac{1-\frac{3\,\GN\,M}{2\,r}}
{\left(1- \frac{2\,\GN\,M}{r}\right)\left[1-\frac{\GN\,M}{2\,r}(4\beta-1)\right]}
\equiv
B(r)
\ .
\label{ar}
\eeq
The solution~\eqref{ar} depends upon just one parameter $\beta$ and the
Minkowski vacuum is recovered for $M \rightarrow 0$.
The horizon radius $r=R$ on the brane is then determined by the algebraic equation $1/B(R)=0$,
and this black hole is either hotter or colder than the Schwarzschild black hole of equal mass
$M$ depending upon the sign of $(\beta - 1)$~\cite{cfabbri}.
For example, assuming $\beta = 5/4$, one finds two solutions equal to the Schwarzschild
radius $R_S = {2\,\GN\,M}$.
Note also that the four-dimensional Kretschmann scalar
$K^{(I)} =  R_{\mu\nu\rho\sigma} R^{\mu\nu\rho\sigma}$
diverges for $r = 0$ and $r = {3\,\GN\,M}/{2}<R_S$.
\subsection{CFM solution II}
\label{3.2}
The second solution for the metric coefficients reads
\begin{eqnarray}
N_{II}(r)
=
1- \frac{2\,\GN\,M}{r}+ \frac{2\,\GN^2\,M^2}{r^2}(\beta-1)
\quad
\text{and}
\quad
 B_{II}(r)
 =
 B(r)
\ .
\label{ar111} 
\end{eqnarray}
In this case the classical radius $R$ for the black hole horizon is given by
$R = R_S$ and $R = R_S\,(\beta-1/4)$. 
Moreover the Kretschmann scalar diverges at each black hole horizon~\cite{kuerten2014},
hence indicating a physical singularity.
\par
A comprehensive analysis of the causal structure and further features on both CFM
solutions can be found in Refs.~\cite{cfabbri,Casadio:2002uv,kuerten2014}.
\section{Bounding the PPN  parameter in the linear regime}
\label{S3}
As previously mentioned, hydrodynamics can be viewed as a low-energy effective 
description for the low-momentum regime of correlation functions.
We are particularly interested in the Green-Kubo formula in the linear response theory,
where an important fluid transport coefficient, the shear viscosity, arises in response
to a perturbation in the fluid stress-energy tensor.
The Kubo formula can further relate the shear viscosity to the absorption
cross section of low-energy gravitons~\cite{Emparan:1997iv}~\footnote{Some equivalent
approaches can be found, e.g.,~in Refs.~\cite{Policastro:2001yc, KSS}.}.
\par
{
Let us consider an action $S$, and introduce a set of sources $J^a$ coupled to a set of
operators ${O^a}$, namely, $S \mapsto S + \int d^4x\,J_a (x)\,{O^a}(x)$, where
$x^\mu=(x^0,x^i)$. 
If the space-time is flat and the vacuum expectation values
(VEV) of all the ${O^a}$ vanish in the absence of the sources $J_a$, one has
\be 
\langle{O^a(x)}\rangle
=
-\int dy\,G^{a|b}_R(x ; y)\, J_b(y)
\ ,
\label{Oa}
\ee
where $G_R(x;y)= G_R(x-y)$ is the retarded Green function of ${O^a}$ in position space 
given by
\beq
i\,G^{a|b}_R(x ; y)
=
\theta(x^0 - y^0)\langle [{O}^a(x), {O}^b(y)]\rangle
\ .
\eeq
Using the interaction picture in QFT~\cite{Natsuume:2014sfa},
one can immediately see that the variation 
\beq
\label{deltao}
\delta\langle{O^a(q)}\rangle
=
-G^{a|b}_R(q)\,J_b(q)
\ ,
\eeq
where $G_R(q)$ denotes the retarded Green function in momentum space with
$q^\mu = (\omega,k)$.
This simple linear formalism can be generalised to curved space-times
and non-vanishing VEV in order to describe a fluid in the conditions of interest
here.
}
\par
A common way to derive the Kubo formula is, in fact, to couple
four-dimensional gravity $g_{\mu\nu}$ to the fluid and determine how the fluid energy-momentum
tensor  responds to gravitational perturbations~\footnote{Note
that, although the space-time is flat in our laboratories, the effect of space-time curvature
on hydrodynamics becomes extremely relevant in astrophysics.}.
In particular, for small metric perturbations which vary slowly in space and time,
$g_{\mu\nu} = \bar g_{\mu\nu} + h_{\mu\nu}$ with $\|h_{\mu\nu}\|\ll1$, the stress tensor can be
doubly expanded in the metric fluctuation $h_{\mu\nu}$ and in gradients~\cite{arnold}.
{This procedure yields a generalisation of Eq.~\eqref{Oa} to higher orders
for the operator $T_{\mu\nu}$, namely}
\beq
\label{exp123}
\langle T^{\mu\nu}(x)\rangle
&=&
\langle T^{\mu\nu}\rangle_{h=0}
-\tfrac 12
\int d^4 y \,G^{\mu\nu| \rho\sigma}_{\rm R}(x;y)\,h_{\rho\sigma}(y)
\nonumber
\\
&&
+
\tfrac 18 \int d^4 y\,\int d^4z\,
G^{\mu\nu| \rho\sigma| \tau\zeta}_{\rm R}(x;y,z)\,
h_{\rho\sigma}(y)\, h_{\tau\zeta}(z)
+\dots
\nonumber
\\
&\equiv&
\langle T^{\mu\nu}_{(0)}\rangle
+\langle T^{\mu\nu}_{(1)}\rangle
+\langle T^{\mu\nu}_{(2)}\rangle
+\cdots
\ ,
\eeq
where $G^{\mu\nu|\dots}_{R}$ are retarded $n$-point correlators,
with the measurement point $z$ having the largest time, and we also assume
the metric fluctuations $h_{\rho\sigma}$ vanish in the far past. The fluid response is then obtained 
from the stress tensor conservation law $\nabla_\mu T^{\mu\nu}=0$, together with the
condition that the fluid describes a conformal theory, namely, $T^\mu_{\;\;\mu}=0$. 
We shall here analyse the first-order formalism, and illustrate how our results
can be further refined by considering second-order terms in the next section.
\par
At zero order in derivatives, the energy momentum tensor reads
\beq
\label{T0}
T^{\mu\nu}_{(0)}
=
(\epsilon+P)\,u^\mu\, u^\nu+P\,\bar g^{\mu\nu}
\ ,
\eeq
where $u^\mu$ is the fluid four-velocity, $\epsilon$ denotes its energy density, $P$ the pressure
and $\bar g_{\mu\nu}$ is again the unperturbed metric on the four-dimensional boundary where the
fluid lives~\footnote{Minkowski space-time is just a particular case corresponding to
a Schwarzschild-type black brane in the bulk.}.

Since in the linear response theory the metric can be seen as the source of the
stress-energy tensor, the response of the two-point function of the associated
perturbed metric can then be written as the first order term $\langle T_{\mu\nu}^{(1)}\rangle$
in the expansion~\eqref{exp123}~\cite{sonn}, 
\beq
\label{tmunu}
\langle T^{\mu\nu}(x)\rangle
\sim
\int dy\,G^{\mu\nu|\alpha\beta}_R(x;y)\,h_{\alpha\beta}(y)
\ ,
\eeq
where the retarded Green function is $G_R^{\mu\nu|\alpha\beta}=\langle T^{\mu\nu}(x)\,T^{\alpha\beta}(y)\rangle$.
In the gauge/gravity duality, the metric is dual to the stress-energy
tensor~\cite{Aharony:1999ti}.
Hence, from the  gravity point of view of the correspondence, the above perturbation
should arise from metric fluctuations $h_{\mu \nu}$ of the appropriate black brane metric. 
\par
{
The first-order expansion (in derivatives) of the hydrodynamical stress-energy tensor,
also known as the constitutive equation, includes dissipative terms, as well as shear and bulk viscosities.
In curved space-times, the first order contribution is given by~\cite{Son:2007vk}
\beq
\label{taumunu}
T^{\mu\nu}_{(1)}
=
-P^{\mu\alpha}P^{\nu\beta}\left[\eta\left(\nabla_\alpha u_\beta
+\nabla_\beta u_\alpha
- \frac23\, \bar g_{\alpha\beta}\,\nabla_\lambda u^\lambda\right)
+\zeta \,\bar  g_{\alpha\beta}\nabla_\lambda u^\lambda\right]
\ ,
\eeq
where $\eta$ is the shear viscosity, $\zeta$ the bulk viscosity,
$\nabla_\mu$ represents the covariant derivative in the (generally curved)
space-time metric $\bar g_{\mu\nu}$.
In addition, $P^{\mu\nu} = \bar g^{\mu\nu}+u^\mu u^\nu$ 
is the projection tensor along spatial directions, that enables us to express the
constitutive equation in a covariant manner.
}
\par
{
We shall next restrict ourselves to a particular type of perturbations which
is simpler to treat using the AdS/CFT correspondence. 
In fact, we suppose, as usual, that the contribution to the shear viscosity $\eta$
comes only from the component $h_{12} = h_{12}(t)$ of $h_{\mu\nu}$,
with all other $h_{\mu\nu} = 0$~\cite{Natsuume:2014sfa,sonn}. 
Since we assumed such fluctuations around thermal equilibrium are small,
we can say the fluid has uniform temperature ${\mathcal T}(x^\mu)={\mathcal  T}_0$ and is at
rest in the chosen frame, that is $u^\mu = (1,u^i=0)$.
Moreover, this perturbation has spin-2 with respect to the spatial $SO(3)$
group and it cannot excite linear order fluctuations of the velocity
(which is a vector) or of the temperature (a scalar).
Therefore, $u^i=0$ and ${\mathcal T} ={\mathcal T}_0$ remain valid up to linear order.
 }
\par
The shear viscosity of the dual theory can be computed from gravity 
in a number of equivalent approaches~\cite{Policastro:2001yc, KSS,Natsuume:2014sfa}.
We shall employ the Kubo formula, which relates the viscosity to equilibrium correlation functions,
in order to compute the $T^{12}$ component of the stress-energy tensor, in a large distance
and long time scale regime.
The non-vanishing contribution in the covariant derivative arises from the Christoffel symbol
\beq
\label{chris}
\nabla_1u_2
=
\partial_1 u_2-\Gamma^\alpha_{12}u_\alpha
=
-\Gamma_{12}^0\,u_0
=
-\frac{1}{2}\,\partial_0 h_{12}
\ .
\eeq
{The component $\nabla_2u_1$ is obtained analogously
and the other components vanish.
Hence, only the first two terms inside the brackets in Eq.~\eqref{taumunu}
contribute to $T^{\mu\nu}_{(1)}$, and we have}
\beq
\label{1212}
{}{
\delta\langle T_{(1){12}}\rangle
\sim
-\eta\left(\nabla_1u_2+\nabla_2u_1\right)
=
-\eta\,\partial_0 h_{12}
\ .}
\eeq
{}{By taking the Fourier transform of Eq.~\eqref{1212},
one obtains~\cite{sonn}}
\beq
\label{1122}
{}{
\delta\langle T_{(1){12}}(\omega, k=0)\rangle
=
i\,\omega\,\eta\, h_{12}
\ .}
\eeq
{}{A perturbed fluid Lagrangian is correspondingly given by
$\delta\mathcal{L} = h_{\mu\nu}(x^0)\,T^{\mu\nu}(x^\alpha) = h_{12}(x^0)\,T^{12}(x^\alpha)$,
for which Eq.~\eqref{deltao} reads}
\beq
\label{txyy}
{}{
\delta\langle T^{12}\rangle
=
-G_R^{12|12}(q)\,h_{12}
\ ,}
\eeq
{}{where $G_R^{12|12}(q) = -i\int d^4 x\, e^{-i\,q_\mu\, x^\mu}\,\theta(x^0)\,
\langle T^{12}(x^\mu)\,T^{12}(0)\rangle$.
Eq.~\eqref{1122} represents the same linear response relation as Eq.~\eqref{txyy},
and the Green-Kubo formula can therefore be obtained as~\cite{Natsuume:2014sfa,sonn}}
\beq
\label{GK}
\eta
=
-\lim_{\omega\to0}\frac{\Im\, G_R^{12|12}(\omega,0)}{\omega}
\ ,
\eeq
where $\Im$ denotes the imaginary part.
\par
Our first concern here is to provide analytic expressions for asymptotically AdS black strings
which solve the full five-dimensional Einstein equations.
The CFM black strings are ``tubular'', codimension one, black branes in AdS \cite{kuerten2014},
corresponding to a late time behaviour of their dual fluid.
Two families of such solutions are obtained from the same assumption that leads to the CFM
brane metrics reviewed in Section~\ref{S2},
namely by  relaxing the condition that the metric coefficient $g_{tt}$ is (minus) the inverse of the
metric coefficient $g_{rr}$.
These two families of black strings are likewise expressed in terms of the PPN parameter
$\beta$~\cite{cfabbri}, and reduce to the five-dimensional Schwarzschild black string solution, 
when $\beta\to1$. {CFM black branes in AdS can be also obtained as a perturbative approach of CFM black holes on the brane \cite{kuerten2014}.}
Our five-dimensional metrics are, in particular, given by
\be
\label{eq09}
g_{AB}\,dx^{A}\, dx^{B}
=
-N(r)\,dt^2+B(r)\,dr^2 + r^2\,d\Omega^2
\ ,
\ee
where $ d\Omega^{2}={\ell^{-2}}\,d\vec{x}^2 ={\ell^{-2}}( dx_{1} ^{2} +dx_{2} ^{2} + dx_{3}^{2})$,
the length $\ell$ is related to the AdS curvature~\cite{sadeghi,HoffdaSilva:2012em} and the metric
coefficients $N$ and $B$ equal the CFM expressions~\eqref{ar} or \eqref{ar111}.
The entropy of these black strings can be computed by using the Hawking-Bekenstein
formula~\cite{KSS},
\be
S
=
\frac{A}{4\,G_5}
=
\frac{R^3\,V_3}{4\,G_5\,\ell^3}
\ ,
\ee
where $R$ is the horizon radius determined by $1/B(R)=0$ and $V_3=\ell\,\int d\Omega$.
Moreover, the volume density of entropy $s=S/V_{3}$ must be finite, which requires that
$\beta <5/4$ in Eqs.~\eqref{ar} and \eqref{ar111}.
It is also convenient to trade the radial coordinate $r$ for the dimensionless $u={R }/{r}$,
and define the length $\bar M = \GN\,M$, so that 
\be
\label{eq22}
B(u)
=
\frac{1-\frac{3\,\bar M}{2\,R}\,u}{\left(1-\frac{2\,\bar M}{R}\,u\right)\left[1-\frac{\bar M}{2\,R}\,u\,(4\beta-1)\right]}
\ee
and the metric finally reads
\be
\label{eq23}
g_{AB}\,dx^{A}\,dx^{B}
=
-N(u)\,dt^{2} +B(u)\,\frac{R^{2}}{u^4}\, du^{2}  +\frac{R^{2} }{u^2}\, d\Omega^2 
\equiv
g_{uu}\, du^{2} +g_{\mu \nu }\, dx^{\mu }\, dx^{\nu }
\ .
\ee
\par
Our main aim is now to predict a value for $\beta$ exclusively from
the KSS bound~\eqref{kss}, implied by the fluid/gravity correspondence.
The background metric~\eqref{eq23} can be perturbed as
$g_{AB} \mapsto g_{AB} + h_{AB}$~\cite{Policastro:2001yc,KSS0}.
As mentioned before, the contribution to the shear viscosity comes from a single
component of the metric fluctuations.
Denoting this term by $\phi=\phi(t,u,\vec x)$, the corresponding wave equation
reads~\cite{KSS,sonn}
\be
\label{eq26}
\partial _{u} \left(\sqrt{-g}\, g^{uu}\, \partial _{u} \phi \right)
+\sqrt{-g}\,g^{\mu \nu }\, \partial_{\mu } \partial_{\nu } \phi
=
0
\ .
\ee
By taking the Fourier transform along the time direction and disregarding the spatial momentum, 
$\phi\simeq e^{i\,\omega\,t}\,\Phi(u)$, one finds
\be
\label{eq27}
\partial _{u} \left(\sqrt{-g}\, g^{uu} \partial _{u} \Phi \right)
-{\sqrt{-g} }\,g^{tt}\, \omega^{2}\, \Phi
\simeq
0
\ ,
\ee
which can be specialised for both CFM metrics I and II, respectively provided by Eqs.~\eqref{ar}
and \eqref{ar111}, that is
\be
\label{eq281}
\frac{d^{2} \Phi}{du^{2} }+\frac{V}{u}\frac{d\Phi}{du}
+\left(1-\frac{2\,\bar M}{R}\,u\right)\omega^2\,\Phi
=
0
\ ,
\ee
where $V=V_{I,II}$, with 
\be
\label{eq281}
V_I
=
{R \left[\frac{1}{(4 \beta -1) \,\bar M\, u-2\, R}+\frac{1}{4\,\bar  M\, u-2\, R}+\frac{1}{2\, R-3\,\bar  M\, u}\right]
+\frac{5}{2}}
\ ,
\ee
for the CFM case I and 
\be
\label{eq282}
V_{II}
=
{R \left[\frac{1}{(4\, \beta -1)\, \bar M\, u-3\, R}+\frac{1}{4\,\bar  M\, u-2\, R\,\beta}
+\frac{1}{4\,R-3\, \bar M\, u}\right]
+\frac{9}{2}}
\ ,
\ee
for the CFM case II.
Following the gauge/gravity prescription, we can then proceed to calculate the retarded Green
function~\cite{Policastro:2001yc,poli2,KSS},
\be
\label{eq37}
G_R (\omega,\vec 0;\beta)
=
\left.-\sqrt{-g}\, g^{uu} \, \Phi^{*}\,\frac{d \Phi}{du}\right|_{u \to 0}
\ ,
\ee
for both cases in Eqs.~\eqref{eq281} and \eqref{eq282}, and then derive the shear viscosity
from the Green-Kubo formula~\eqref{GK}.
A bound on the PPN parameter $\beta$ can finally be obtained from the KSS
lower bound~\eqref{kss}.
{From the propagator~\eqref{eq37}, by inserting Eq.~\eqref{GK} in Eq.~\eqref{kssm},
one in fact finds}
\be
\label{dfer}
\frac{\eta(\beta)}{s(\beta)}
=
-\frac{1}{s(\beta)}\,
{\mathop{\lim }\limits_{\omega \to 0} \frac{\Im\, G_R (\omega, k=0;\beta)}{\omega } }
=
\frac{1}{4\,\pi}
\ ,
\ee
where we remarked that both $s$ and $\eta$ do depend on $\beta$.
However, the solutions to Eq.~\eqref{eq37} are cumbersome linear combinations of integral
exponentials and Hermite functions, and we shall therefore just display the results graphically, 
 {in the near horizon approximation}.
\par
\begin{figure}[t]
\begin{center}
\includegraphics[width=2.5in]{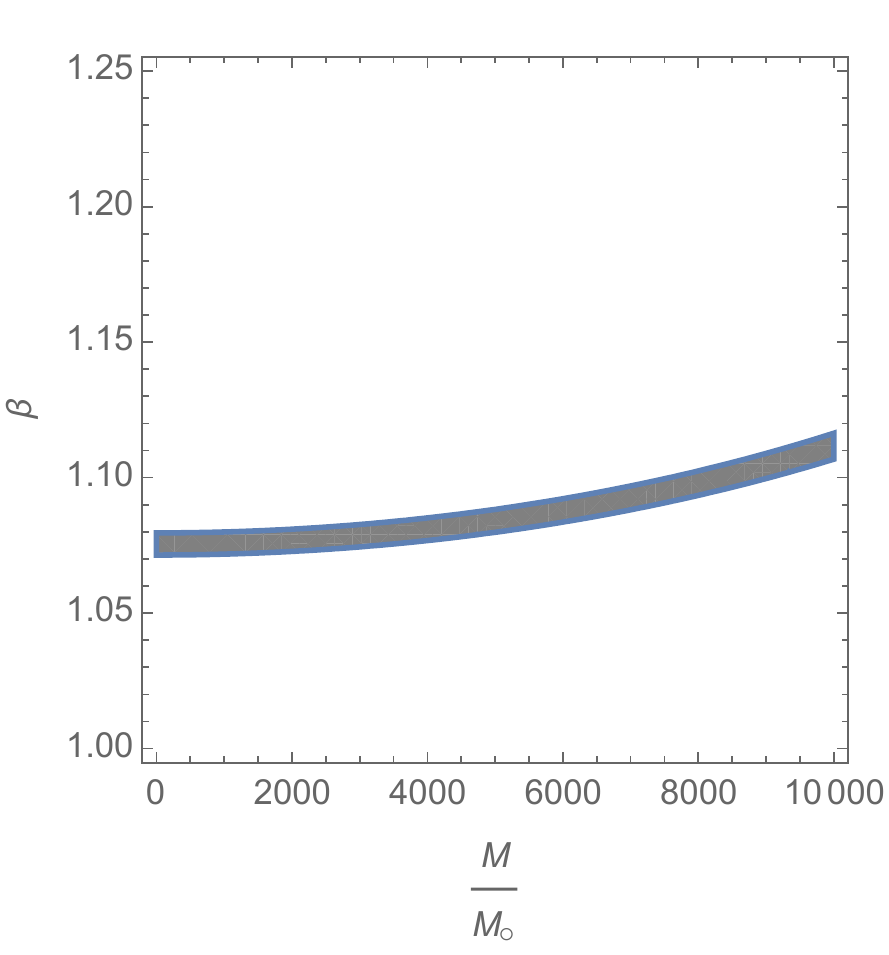}
\qquad 
\includegraphics[width=2.5in]{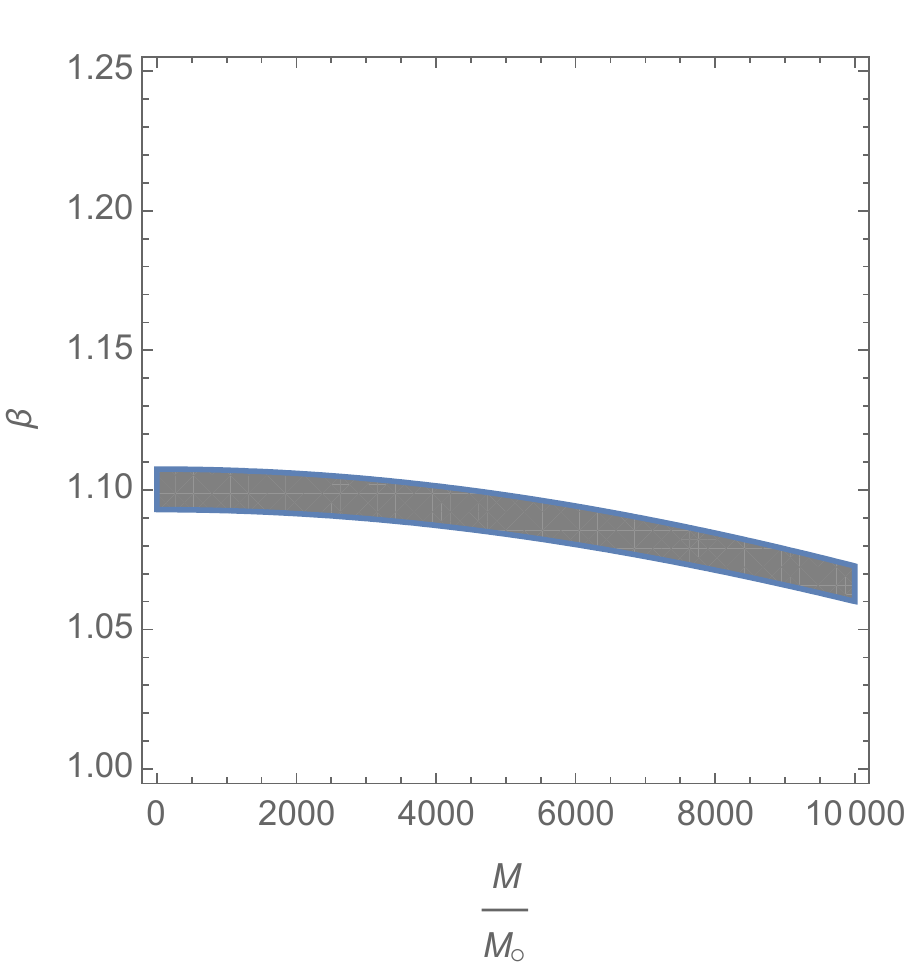}
\end{center}
\caption{Allowed range for the PPN parameter $\beta$ as a function of the mass $M$
(in Solar mass units), for CFM~I (left panel) and CFM~II (right panel) in the linear regime.
\label{fig1}
}
\end{figure}
The resulting bound for $\beta$ is displayed graphically in Fig.~\ref{fig1}, where the grey bands
correspond to the allowed range for varying black string mass $M$ in units of the solar
mass $M_\circ$.
{The plot in the left panel is for the CFM solution I and $M_\circ<M<10^4\,M_\circ$.
For $M=M_\circ$, the PPN parameter must lie in $1.0772\lesssim \beta \lesssim1.0803$, 
whereas $1.1074\lesssim\beta \lesssim 1.1143$ for a brane-world CFM~I black hole having a mass
equivalent to $10^4$ solar masses.
Similarly, for the CFM~II black hole, we point out that if $M=M_\circ$, the bound on the shear
viscosity-to-entropy ratio yields $1.0913\lesssim\beta \lesssim 1.1093$, and
$1.0602\lesssim\beta\lesssim 1.0731$ for $M=10^4\,M_\circ$. }
On the other hand, the perihelion shift provides an observational value of $|\beta-1| \lesssim 0.003$
and the study of the Nordtvedt effect yields the bound $|\beta-1|\lesssim 0.00023$~\cite{will}.
Both experimental bounds fall outside the bands shown in Fig.~\ref{fig1},
although they are close in value.
Nevertheless, we shall argue in the next Section how the situation can be significantly improved.  

\section{Second order improvements}
\label{S4}
The results obtained in Section~\ref{S3} are not far from the observational data, nevertheless they
still appear to be ruled out by the experimental bounds. 
This could signal that the fluid/gravity correspondence fails for these systems, or that the approximations
we employed are too strong.
In particular, Eq.~\eqref{dfer} was obtained in the linear hydrodynamics regime, 
 {corresponding to the first two terms in Eq.~\eqref{exp123}, that is Eqs.~\eqref{T0}
and~\eqref{taumunu}.
We can instead tackle the non-linear dynamics by keeping higher-order terms in the expansion~\eqref{exp123}
of the stress-energy tensor in velocity and gradients multiplied by the corresponding hydrodynamic
coefficients~\cite{hubeny}.}
In particular, we shall here include the second-order regime of a dual hydrodynamic
prescription~\cite{Bu:2014ena,hubeny,Kovtun:2011np,Moore:2012tc,arnold}, in order to refine the bounds on the PPN
parameter $\beta$.
\par
{}{
The second-order hydrodynamic analysis is rather
involved~\cite{hubeny,Bu:2014ena,Kovtun:2011np,arnold,Grozdanov:2014kva,nastase}, so let us first introduce some
compact notation.
For any second rank tensor $O^{\alpha\beta}$, we denote
$O^{\langle \alpha\beta\rangle}\equiv \frac12 P^{\alpha\mu}P^{\beta\nu}O_{(\mu\nu)}-\frac13P^{\alpha\beta}P^{\mu\nu}O_{\mu\nu}$,
which is transverse to the fluid motion, $u_\mu \,O^{\<\mu\nu\>} = 0$, and traceless $g_{\mu\nu}\, O^{\<\mu\nu\>}=0$. 
In addition, 
$\sigma^{\alpha\beta} = P^{\alpha\mu}P^{\beta\nu}\nabla_{(\mu} u_{\nu)}-\frac23  P^{\alpha\beta}P^{\mu\nu}\nabla_\mu u_\nu$
is the shear tensor and $\Omega^{\alpha\beta} = P^{\alpha\mu}P^{\beta\nu}\nabla_{[\mu} u_{\nu]}$ denotes the vorticity
tensor, as they appear in the $T^{\mu\nu}_{(1)}$ of Eq.~\eqref{taumunu}. 
We next express derivatives along the fluid $4$-velocity in terms of transverse
derivatives by means of the zero-order equations of motion~\cite{arnold},}
{
\beq
\label{termosall}
u^\rho\,\nabla_\rho u^\mu
=
- \nabla_\perp^\mu \ln T\ ,
\qquad
u^\rho\,\nabla_\rho \ln T
=
- \frac12\, \left(\nabla_\perp\! \cdot u\right)
\ ,
\eeq
where $\nabla^\mu_\perp \equiv P^{\mu\alpha}\nabla_\alpha$ and
$\nabla_\perp\!\cdot u = \nabla\cdot u\equiv \nabla_\mu u^\mu$.
All possible contributions to the stress-energy tensor are 
$\sigma^{\mu\nu}(\nabla{\cdot}u)$, 
${\sigma^{\<\mu}}_\lambda \sigma^{\nu\>\lambda}$, $\nabla^{\<\mu}\ln T\, \nabla^{\nu\>}\ln T$,
${\sigma^{\<\mu}}_\lambda \Omega^{\nu\>\lambda}$,  $\nabla^{\<\mu}\nabla^{\nu\>} \ln T$, 
$u_\alpha R^{\alpha\<\mu\nu\>\beta} u_\beta$, $ {\Omega^{\<\mu}}_\lambda \Omega^{\nu\>\lambda}$, 
and $R^{\<\mu\nu\>}$, 
where $R^{\alpha\beta\gamma\delta}$ is the Riemann tensor.
But the only ones that transform homogeneously under Weyl transformations
are~\cite{arnold}
\begin{align}
&
\op{1}
=
R^{\<\mu\nu\>} - \left(\nabla^{\<\mu} \nabla^{\nu\>} \ln T 
- \nabla^{\<\mu} \ln T\, \nabla^{\nu\>} \ln T\right),
\qquad
\op{2}= R^{\<\mu\nu\>}-u_\alpha u_\beta R^{\alpha\<\mu\nu\>\beta}
\ ,
\label{Rmunu}
\\
&
\op{3}
=
{\sigma^{\<\mu}}_\lambda \sigma^{\nu\>\lambda}
\ ,
\qquad
\op{4}
=
{\sigma^{\<\mu}}_\lambda \Omega^{\nu\>\lambda}
\ ,
\qquad
\op{5}
=
{\Omega^{\<\mu}}_\lambda \Omega^{\nu\>\lambda}
\ .
\end{align}
Eq.~(\ref{Rmunu}) can be recast as~\cite{arnold}
$ u^{\<\rho}\,\nabla_\rho \sigma^{\mu\nu\>}+ \frac 12\sigma^{\mu\nu}(\nabla{\cdot} u)$, 
which by using (\ref{termosall}), reduces to the linear combination  
$\op{1}-\op{2}-(1/2)\,\op{3}-2\,\op{5}$.
The second order dissipative part in the expansion of the stress-energy tensor~\eqref{exp123}
must therefore read
\beq
\label{pmunu2}
T^{\mu\nu}_{(2)}
&=&
\eta\,\tau_\Pi \left[ u^{\<\rho}\nabla_\rho\sigma^{\mu\nu\>} + \frac12\sigma^{\mu\nu}
    (\nabla{\cdot}u) \right]
+\kappa
\left(R^{\langle\mu\nu\rangle} - 2u_\rho u_\tau R^{\rho\langle\mu\nu\rangle\tau}\right)
\nonumber
\\
&&
+ \lambda_1\, \sigma^{\<\mu}_\tau\,\sigma^{\nu\>\tau}
+\lambda_2\, \sigma^{\<\mu}_\tau\,\Omega^{\nu\>\tau}
+\lambda_3\, \Omega^{\<\mu}_\tau\,\Omega^{\nu\>\tau}
\ ,
\eeq 
where the $\kappa$ term contributes to the two-point Green's function of the
stress-energy tensor. }
The $\lambda_{i}$ terms are non-linear in velocity and the parameter $\tau_\Pi$
can be thought of as the relaxation time~\cite{arnold}. 
It is worth mentioning that the second-order hydrodynamic transport coefficients $\kappa$ and $\lambda_i$
have been calculated from the leading order AdS/CFT correlators at finite chemical potential and in the presence
of fundamental matter~\cite{Moore:2012tc,arnold,Grozdanov:2014kva}. They correspond to the black brane horizon response properties. 
Moreover, these coefficients have been derived for strongly coupled plasma described holographically
by an Einstein-scalar action in the bulk, as functions of the temperature in lattice QCD
thermodynamics~\cite{Finazzo:2014cna}.
{Nonetheless, there is not a final word concerning the values of these parameters.
For conformal theories at finite temperature, and zero chemical potential, they
scale with suitable powers of the temperature and may depend
on coupling constants and eventually on the rank of the gauge group~\cite{arnold,Nakayama:2012vs}.
In fact, there is an intrinsic ambiguity in second order viscosity parameters
in relativistic hydrodynamics~\cite{Nakayama:2012vs}}.
 {For example, one of the coefficients can be calculated as
$ \lambda_3= -4 \lim_{\substack{k_1\to0\\k_2\to0}}
\frac{\partial}{\partial_{k_1}} \frac{\partial}{\partial_{k_2}}
\lim_{\substack{\omega_1\to0\\\omega_2\to0}}G^{xy| 0x| 0y}$.} 
 {The expressions thus obtained are very similar to those given}
in Refs.~\cite{arnold, Grozdanov:2014kva}, namely
\begin{subequations}
\begin{eqnarray}
\kappa
&=&
\frac{N_c^2\,{\mathcal  T}^2}{8}\left(1-10\,\gamma\right)
\label{trc1}
\\
\tau_\Pi
&=&
\frac{2-\ln 2}{2\pi {\mathcal  T}}+\frac{375\gamma}{4\pi {\mathcal T}}
+\ldots
\label{trc12}
\\
\lambda_1
&=&
\frac{N_c^2\,{\mathcal  T}^2}{16}\left(1+350\,\gamma\right)
\label{trc2}
\\
\lambda_2
&=&
-\frac{N_c^2\,{\mathcal  T}^2}{16}\left[2\,\ln 2 + 5\, (97 + 54\,\ln 2)\, \gamma + \ldots\right]
\label{trc3}
\\
\lambda_3
&=&
\frac{25\,N_c^2\,{\mathcal  T}^2}{2}\,\gamma+\ldots
\label{trc4}
\end{eqnarray}
\end{subequations}
where $\gamma = (g^2\,N_c)^{-3/2}\,\zeta(3)/8$ and ${\mathcal  T}$ again denotes the system
temperature.
We naturally choose the latter to equal the black string Hawking temperature, given by Eqs.~(13)
and~(21) in Ref.~\cite{cfabbri} for the CFM solutions I and II respectively, and depending
on $M$ and $\beta$.
{The strong coupling limit $\gamma\to0$ can be also
obtained~\cite{Policastro:2001yc,hubeny}.
We emphasise that our results below are independent of changes in the above
transport coefficients, at least for the parameters derived in Refs.~\cite{Policastro:2001yc,hubeny,Nakayama:2012vs}.} 
\par
{Similar steps to the ones yielding Eq.~\eqref{chris} and the Kubo formula~\eqref{GK}
can be repeated.
If one now takes, as usual in the literature, $h_{12}=h_{12}(t,z)$, Eq.~\eqref{pmunu2} yields}
\beq
T_{(1){12}}+T_{(2)12}
=
-P\, h_{12} -\eta\, \dot h_{12} + \eta\,\tau_\Pi\, \ddot h_{12}
-\frac\kappa 2 \left[\ddot h_{12} + \d_z^2 h_{12}\right]
\ ,
\eeq
where dots denote time derivatives. 
The retarded Green  function then reads 
\beq
G_R^{12|12}(\omega, k)
=
P - i\,\eta\,\omega + \eta\,\tau_\Pi\, \omega^2-\frac\kappa 2 \left[\omega^2+k^2\right]
\ ,
\eeq
which is the analogue of the Kubo formula with second-order derivative terms taken into account.
We then see the kinetic parameters $\tau_\Pi$ and $\kappa$ are the coefficients of the $\omega^2$
and $k^2$ terms in the low-momentum expansion of the retarded propagator.
\par
\begin{figure}[t]
\begin{center}
\includegraphics[width=2.49in]{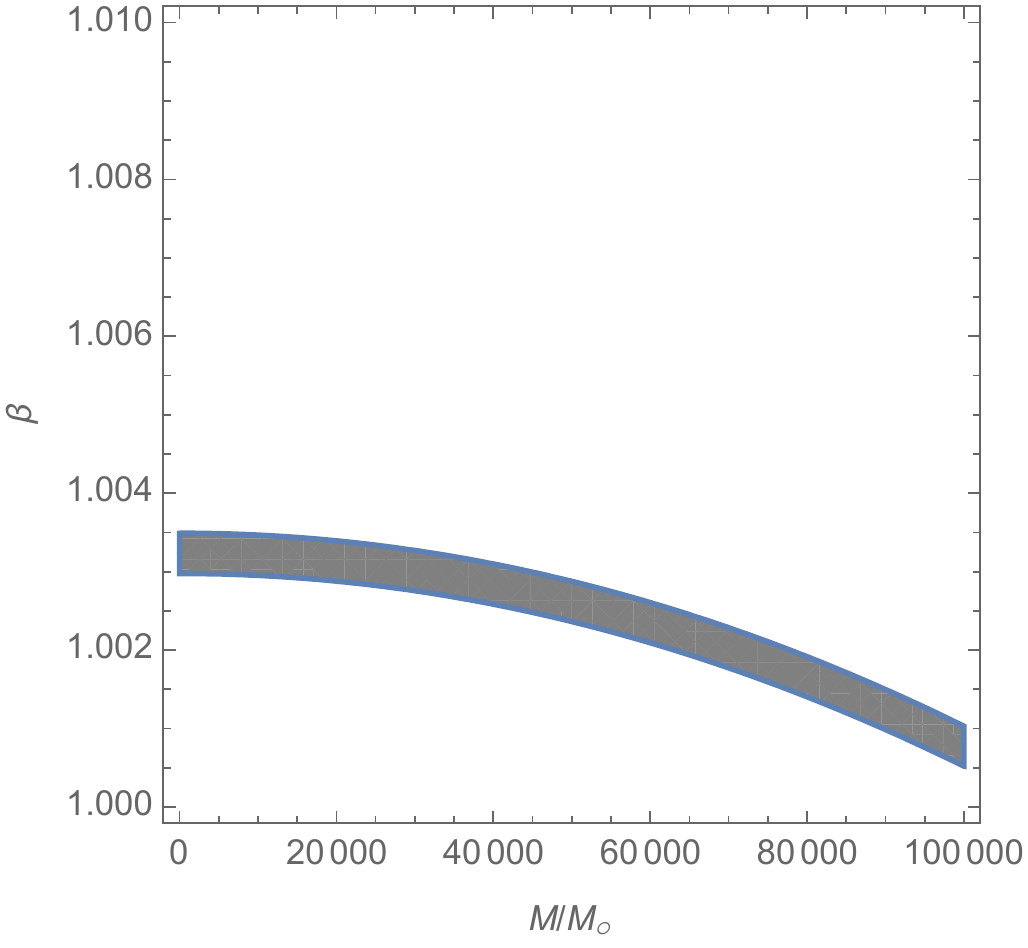}\qquad 
\includegraphics[width=2.5in]{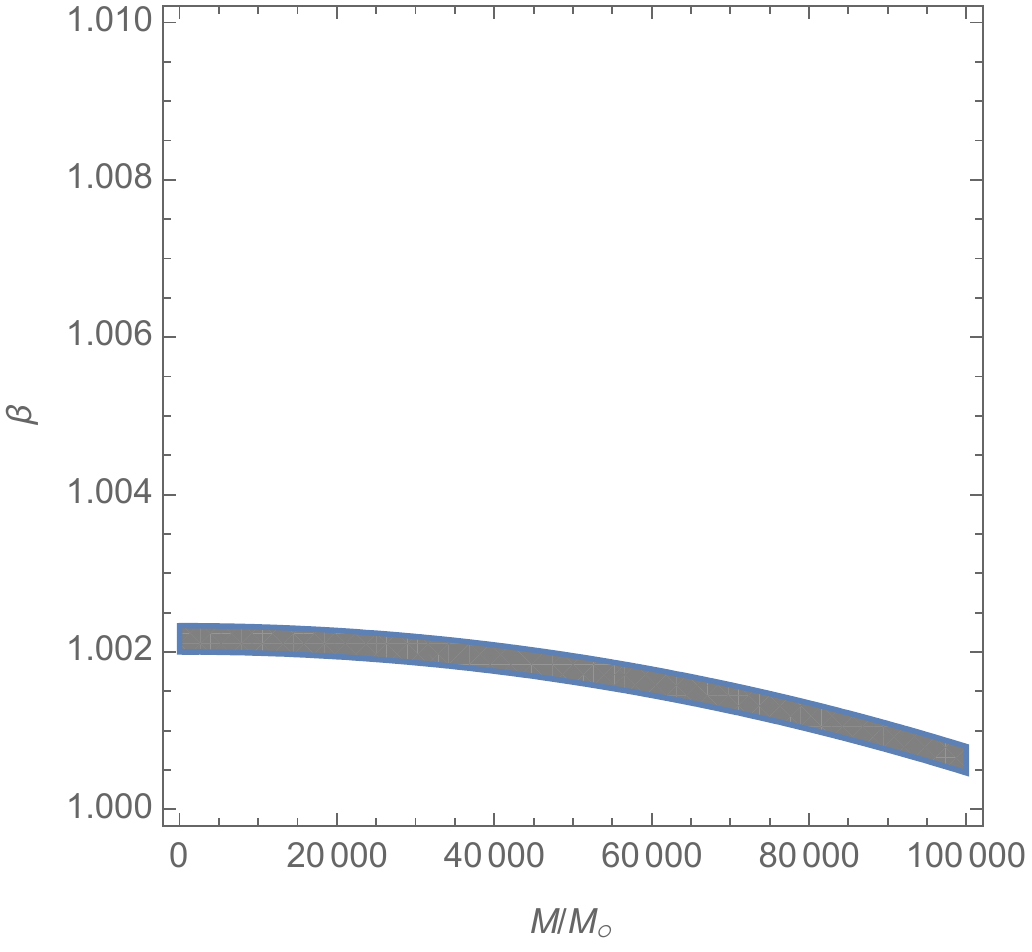}
\end{center}
\caption{Allowed range for the PPN parameter $\beta$ as a function of the mass $M$
(in Solar mass units), for CFM~I (left panel) and CFM~II (right panel) including second-order
corrections.}
\label{fig2}
\end{figure}
Upon employing the above expressions and repeating the analysis of Sect.~\ref{S3}, one finds
new bounds for the PPN parameter $\beta$, and preliminary results are shown in Fig.~\ref{fig2}. 
All terms in Eq.~\eqref{pmunu2} are included, with transport coefficients provided by
Eqs.~\eqref{trc1}-\eqref{trc4} correspond to the second term in the right hand side of
Eq.~\eqref{nccc}.
{Clearly, the strong coupling limit $\gamma\to0$ can be taken and
Eq.~\eqref{kss} then appears as a particular case of Eq.~\eqref{nccc}.} 
The plot in the left panel is for the CFM solution I and $M_\circ<M<10^5\,M_\circ$.
For $M=M_\circ$, the PPN parameter must lie in $1.0021\lesssim \beta \lesssim1.0042$, 
whereas $1.0001\lesssim\beta \lesssim 1.0013$ for a brane-world CFM~I black hole having a mass
equivalent to $10^5$ solar masses.
Similarly, for the CFM case~II, it is worth to mention that when $M=M_\circ$, the bound on the shear
viscosity-to-entropy ratio yields $1.0014\lesssim\beta \lesssim 1.0042$, and
$1.0019\lesssim\beta\lesssim 1.0044$ for $M=10^5\,M_\circ$. 
The numerical stability of these results has been checked by considering several different orders of
magnitude for the t'Hooft coupling, and the differences were always found to be negligibly small.
Moreover, refined expressions involving more transport coefficients, like in Ref.~\cite{Finazzo:2014cna},
could be used.
However, such contributions in the viscosity-to-entropy ratio would amount at most to order $10^{-5}$
corrections in the bounds for $\beta$.
Similarly, we tested other possible choices for the second-order viscosity parameters~\eqref{trc1}-\eqref{trc4}
and the differences with respect to our results were always found to be negligibly small.
 {A comment is due about the relaxation time $\tau_\Pi$ in Eq.~\eqref{trc12},
which turns out to affect the viscosity-to-entropy ratio by a negligible order of $10^{-6}$.
This transport coefficient, which appears in the first term on the r.h.s.~of Eq.~\eqref{pmunu2}, 
varies with the inverse of the temperature, unlike the other coefficients in Eqs.~\eqref{trc1}
and~\eqref{trc2}-\eqref{trc4} which vary with the square of the temperature.
Given the smallness of the black string temperature \cite{Abdalla:2006qj}, it might seem that it would affect our results.
However, this is not the case, since $\tau_\Pi$ multiplies a term that contains
third-order covariant derivatives of the fluid velocity, and, by Eq.~\eqref{chris}, is proportional
to the third-order derivative of the perturbation $h_{12}$ which, in the regime of dual fluids
under perturbations in the fluid/gravity setup, is identically zero~\footnote{It would influence
regimes of shock waves in fluids~\cite{hubeny,Horowitz:1999gf}.}.}
These preliminary results including second-order terms are thus fully compatible with the observational
bound $|\beta-1| \lesssim 0.003$ provided by the perihelion shift~\cite{will}.
Our aim here was to show the fluid/gravity correspondence can be a powerful method,
{by applying the KSS result.
In particular, this correspondence allows one to obtain reasonable theoretical bounds for the already
existing observational values of a post-Newtonian parameter.
It is also worth emphasising that the CFM black brane should not satisfy any
four-dimensional PPN bound \emph{a priori\/}, since the black brane resides in five dimensions.
It is, at least, an intriguing numerical coincidence that the PPN bound for the parameter of the
CFM black brane matches the four-dimensional PPN bounds.
Nevertheless, the main result here is that, since the PPN parameter is bound to be close to the unity
also in five dimensions, the CFM black branes must effectively reduce to their Schwarzschild
counterparts.
In fact, both CFM metrics~\eqref{ar} and \eqref{ar111} reduce to the Schwarzschild metric in the
limit $\beta\to 1$.
Here, we proved that the KSS result forces the CFM black branes to reduce to Schwarzschild
black branes.
For the boundary energy-momentum expansion~\eqref{exp123}, the first order approximation~\eqref{taumunu}
provides $|\beta-1|\lesssim 10^{-1}$, whereas the second order approximation~\eqref{taumunu}
provides $|\beta-1|\lesssim 3\times 10^{-2}$.
When the higher order term $\langle T^{\mu\nu}_{(3)}\rangle$ is included in Eq.~\eqref{exp123},
one finds  $|\beta-1|\lesssim 10^{-4}$, suggesting that $\beta\to1$ when sufficiently high-order terms
$\langle T^{\mu\nu}_{(n)}\rangle $ are taken into account in Eq.~\eqref{exp123}.
This is in agreement with the conjecture of Ref.~\cite{conj}, that the Schwarzschild black brane is the
unique static black hole solution of vacuum gravity with a negative cosmological constant.}
\section{Discussion and outlook}
\label{S5}
In this work, we derived the shear viscosity-to-entropy ratio $\eta/s$ as a function of the PPN
parameter $\beta$ of candidate CFM black strings from the Green-Kubo formula in the context
of the fluid/gravity correspondence.
On assuming the KKS universal bound~\eqref{kss}, we then determined the allowed range of
$\beta$ for different black string mass.
Our results show a possible slight dependence of $\beta$ on the mass in the range
$M_\circ<M<10^4\,M_\circ$ in the linear order hydrodynamic expansion (see Fig.~\ref{fig1}).
The non-linear hydrodynamical regime was just started to be investigated including second order
terms.  
In particular, when the terms displayed in Eq.~\eqref{pmunu2} are included, with the transport
coefficients~\eqref{trc1}-\eqref{trc4}, the theoretical bounds are modified and appear in agreement
the four-dimensional observational bounds (see Fig.~\ref{fig2}).
The discrepancy with experimental data shown in Fig.~\ref{fig1} could thus be simply related to the
approximations employed in order to derive Eq.~\eqref{dfer} from the Green-Kubo formula, for instance,
the near horizon limit and the linearised hydrodynamics regime.
Of course, this will be the subject of further investigations.
{We showed that the KSS result implies that the CFM black branes are, effectively,
Schwarzschild black branes, when sufficiently high-order terms in the boundary energy-momentum
tensor~\eqref{exp123} are taken into account, in agreement with the conjectured unicity of the
Schwarzschild black brane as a static black-hole solution of vacuum gravity with a negative
cosmological constant~\cite{conj}.}
\par
As more possible developments, we recall that higher derivative gravity corrections have been
considered~\cite{shenker,arnold}.
The ratio of shear viscosity-to-entropy density indicates that the KSS bound~\eqref{kss} could then
be violated, and the new lower bound $\eta/s\ge 4/25\,\pi$ was proposed~\cite{Cai:2009zv},
based on causality of the dual field theory.
In this context, it seems that such higher derivative gravity corrections make the CFM solutions
inadequate as a black brane prescription.
Indeed, from the bound $\eta/s\gtrsim 4/25\,\pi$, the PPN parameter of these metrics would be
$\beta \simeq 0.7432$, which contradicts the theoretical bound $\beta\geq1$.
A correction for $\eta/s$ was also given in Ref.~\cite{shenker} in the form of a linear combination
of the coefficients $\alpha_i$ of the scalars $R$, $R_{\mu\nu}R^{\mu\nu}$ and
$R_{\mu\nu\rho\sigma}R^{\mu\nu\rho\sigma}$ in the Gauss-Bonnet contribution to the action.
More precisely, in the Gauss-Bonnet context, $\eta/s\simeq (1-8\,\alpha_3)/{4\,\pi}$ and the KSS
bound is violated for $\alpha_3>0$.
However, since the CFM solutions are obtained with no Gauss-Bonnet terms, one can set
$\alpha_3=0$ and there is no anisotropy, so that the KSS bound~\eqref{dfer} is the only
one that does hold for CFM metrics.
The quark-gluon plasma produced in the ultra-relativistic collision of heavy ions at RHIC
and LHC~\cite{adams} is a strongly coupled fluid that is represented by the anisotropic evolution
of the fluid in the initial stages before isotropisation.
Recently, there has been some interest in modelling this anisotropy at strong
coupling~\cite{diego,Critelli:2014kra} and in studying how  observables may be affected by it.
Some of the studies include the computation of the shear viscosity-to-entropy density
ratio~\cite{rebhan,Megias:2013joa}.
Anisotropic scenarios can be further considered to include such possibilities.

\subsection*{Acknowledgements}
R.C.~is partly supported by INFN grant FLAG.
R.T.C.~is supported by CAPES, PDSE and is grateful to R.C.~for the hospitality.
R.dR.~is grateful to CNPq (grants No. 303293/2015-2, and No.~473326/2013-2),
and to FAPESP (grant No.~2015/10270-0) for partial financial support. RdR thanks to Prof. Jorge Noronha for fruitful discussions.

\begin{thebibliography}{99}
%
%
\bibitem{maartens}
R. Maartens, K. Koyama,
\emph{Brane-world gravity},
Living Rev. Rel. {\bf 13}, 5 (2010).

\bibitem{Antoniadis:1998ig}
  I.~Antoniadis, N.~Arkani-Hamed, S.~Dimopoulos and G.~R.~Dvali,
  \emph{New dimensions at a millimeter to a Fermi and superstrings at a TeV}, 
  Phys.\ Lett.\ B {\bf 436} (1998) 257.

\bibitem{Antoniadis:1990ew}
  I.~Antoniadis, \emph{A Possible new dimension at a few TeV}, 
  Phys.\ Lett.\ B {\bf 246} (1990) 377.

\bibitem{landsberg}
S.~Dimopoulos and G.~L.~Landsberg,
\emph{Black holes at the LHC},
Phys.\ Rev.\ Lett.\  {\bf 87} (2001) 161602.

 \bibitem{kuerten2014} 
  R.~da Rocha, A.~Piloyan, A.~M.~Kuerten, C. H. Coimbra-Araujo,
  \emph{Casadio-Fabbri-Mazzacurati Black Strings and brane-world-induced Quasars Luminosity Corrections},
  Class.\ Quant.\ Grav.\  {\bf 30},  045014 (2013).   


\bibitem{clark}
S.~S.~Seahra, C.~Clarkson and R.~Maartens,
  \emph{Detecting extra dimensions with gravity wave spectroscopy: the black string brane-world}, 
  Phys.\ Rev.\ Lett.\  {\bf 94}, 121302 (2005).

\bibitem{cor2014}
R. Casadio, J. Ovalle, R. da Rocha, {\it 
Black Strings from Minimal Geometric Deformation in a Variable Tension Brane-World}, 
 Class. Quant. Grav. \textbf{30}, 175019 (2014).
%

\bibitem{rr22014}
D. Bazeia, J. M. Hoff da Silva, R. da Rocha, {\it Regular Bulk Solutions and Black Strings from Dynamical brane-worlds with Variable Tension}, 
 Phys. Rev. D 
\textbf{90}, 047902 (2014).

\bibitem{Bu:2014ena}
  Y.~Bu and M.~Lublinsky,
  \emph{Linearized fluid/gravity correspondence: from shear viscosity to all order hydrodynamics}, 
  JHEP {\bf 1411}, 064 (2014).

\bibitem{Maldacena:1997re}
J. M.~Maldacena,
{\it The large N limit of superconformal field theories and supergravity,}
Adv.\ Theor.\ Math.\ Phys.\  {\bf 2}, 231 (1998). 


\bibitem{Gubser:1998bc}
S. S.~Gubser, I.  R.~ Klebanov, and A.M.~Polyakov,
{\it Gauge theory correlators from non-critical string theory,}
Phys.\ Lett.\ B {\bf 428}, 105 (1998).



\bibitem{Aharony:1999ti}
O.~Aharony, S. S.~Gubser, J.M.~Maldacena, H.~Ooguri, and Y.~Oz,
{\it Large N field theories, string theory and gravity,}
Phys.\ Rep.\ {\bf 323}, 183 (2000).

\bibitem{Natsuume:2014sfa}
  M.~Natsuume,
  ``AdS/CFT Duality User Guide,''
  Lect.\ Notes Phys.\  {\bf 903} (2015) 
  [{\tt arXiv:1409.3575 [hep-th]}].

\bibitem{Policastro:2001yc}
G.~Policastro, D. T.~Son, and A. O.~Starinets, 
{\it Shear viscosity of strongly coupled ${\cal N} = 4$ supersymmetric 
Yang-Mills plasma,}
Phys.\ Rev.\ Lett.\  {\bf 87}, 081601 (2001).

\bibitem{poli2} 
  G.~Policastro, D.~T.~Son and A.~O.~Starinets,
 \emph{From AdS / CFT correspondence to hydrodynamics,}
  JHEP {\bf 0209}, 043 (2002).



\bibitem{KSS} P. Kovtun, D.T. Son and A.O. Starinets,
{\it Holography and hydrodynamics: Diffusion on stretched horizons},
JHEP {\bf 0310}, 064 (2003).



\bibitem{KSS0} P.~Kovtun, D.T.~Son and A. O.~Starinets,
{\it Viscosity in strongly interacting quantum field theories from black
hole physics},  Phys.\ Rev.\ Lett.\  {\bf 94}, 111601 (2005).

\bibitem{Buchel:2004di}
A.~Buchel, J. T.~Liu, and A.O.~Starinets,
{\it Coupling constant dependence of the shear viscosity in 
${\cal N} = 4$ supersymmetric Yang-Mills theory,}
Nucl.\ Phys.\ B {\bf 707}, 56 (2005).

\bibitem{Emparan:1997iv}
R.~Emparan,
{\it Absorption of scalars by extended objects,}
Nucl.\ Phys.\ B {\bf 516}, 297 (1998).


\bibitem{Cremonini:2011iq}
  S.~Cremonini,
{\it The shear viscosity to entropy ratio: A status report,}
  Mod.\ Phys.\ Lett.\ B {\bf 25}, 1867 (2011).

\bibitem{Cai:2009zv}
  R. G.~Cai, Z. Y.~Nie, N.~Ohta and Y. W.~Sun,
{\it Shear viscosity from Gauss-Bonnet gravity with a dilaton coupling,}
  Phys.\ Rev.\ D {\bf 79}, 066004 (2009).

\bibitem{Lehner:2010pn}
  L.~Lehner and F.~Pretorius, \emph{Black Strings, Low Viscosity Fluids, and Violation of Cosmic Censorship},    Phys.\ Rev.\ Lett.\  {\bf 105}, 101102 (2010).  

\bibitem{OHara}
K. M.~O'Hara,  S. L.~Hemmer, M. E.~Gehm, S. R.~Granade, and J. E.~Thomas,
{\it Observation of a strongly-interacting degenerate Fermi gas of atoms}, 
Science {\bf 298},  2179 (2002).
%

\bibitem{cherman}
A. Cherman, T. D. Cohen, P. M. Hohler,
\emph{A sticky business: the status of the conjectured viscosity/entropy density bound}, {JHEP} {\bf 0802}, 026 (2008).

 \bibitem{hubeny}
 S.~Bhattacharyya, V.~E.~Hubeny, S.~Minwalla and M.~Rangamani, 
 \emph{Nonlinear Fluid Dynamics from Gravity}, JHEP {\bf 0802}, 045 (2008).

 \bibitem{sadeghi} 
  M.~Sadeghi and S.~Parvizi,
 \emph{Shear Viscosity to Entropy Density for a Black Brane in 5-dimensional Einstein-Yang-Mills Gravity}, Class. Quant. Grav. {\bf 32} (2015).

\bibitem{Casadio:2002uv} 
  R.~Casadio and L.~Mazzacurati,
  \emph{Bulk shape of brane world black holes}, 
  Mod.\ Phys.\ Lett.\ A {\bf 18}, 651 (2003).

 \bibitem{bs1}
  R.~da Rocha, J.~M.~Hoff da Silva, \emph{Black string corrections in variable tension brane-world scenarios},
  Phys.\ Rev.\ D {\bf 85}, 046009 (2012).

\bibitem{cfabbri}
R.~Casadio, A.~Fabbri and L.~Mazzacurati,
 {\it New black holes in the brane world?}, 
Phys.\ Rev.\ D {\bf 65},  084040 (2002). 


\bibitem{will}
C.~M.~Will,
  \emph{The Confrontation between general relativity and experiment}, 
  Living Rev.\ Rel.\  {\bf 9}, 3 (2006).

%
\bibitem{Dadhich} N.{} Dadhich, R.{} Maartens, P.{} Papadopoulos, V.{} Rezania, \emph{Black holes on the brane,} {Phys. Lett. B} {\bf 487}, 1  (2000).

 \bibitem{gregory} R. Gregory and R. Laflamme, \emph{Black strings and p-branes are unstable},  Phys. Rev. Lett. {\bf 70}, 2837 (1993).

\bibitem{arnold} P. Arnold, D. Vaman, C. Wu and W. Xiao, \emph{Second order hydrodynamic coefficients from
3-point stress tensor correlators via AdS/CFT},  JHEP {\bf 1110}, 033 (2011).


\bibitem{sonn} 
  D.~T.~Son,
  \emph{Hydrodynamics and gauge/gravity duality}, 
  Nucl.\ Phys.\ Proc.\ Suppl.\  {\bf 192}, 113 (2009).

\bibitem{Son:2007vk} 
  D.~T.~Son and A.~O.~Starinets,
  \emph{Viscosity, Black Holes, and Quantum Field Theory,} 
  Ann.\ Rev.\ Nucl.\ Part.\ Sci.\  {\bf 57}, 95 (2007). 


\bibitem{HoffdaSilva:2012em} 
  J.~M.~Hoff da Silva and R.~da Rocha,
 \emph{Effective Monopoles within Thick Branes,} 
  Europhys.\ Lett.\  {\bf 100}, 11001 (2012).

\bibitem{Kovtun:2011np} 
  P.~Kovtun, G.~D.~Moore and P.~Romatschke,
  \emph{The stickiness of sound: An absolute lower limit on viscosity and the breakdown of second order relativistic hydrodynamics}, 
  Phys.\ Rev.\ D {\bf 84}, 025006 (2011).


\bibitem{Moore:2012tc} 
  G.~D.~Moore and K.~A.~Sohrabi,
  \emph{Thermodynamical second-order hydrodynamic coefficients}, 
  JHEP {\bf 1211}, 148 (2012).



\bibitem{Grozdanov:2014kva} 
  S.~Grozdanov and A.~O.~Starinets,
 \emph{On the universal identity in second order hydrodynamics}, 
  JHEP {\bf 1503}, 007 (2015).

\bibitem{nastase} H. Nastase, ``Introduction to the AdS/CFT Correspondence'', Cambridge Univ. Press, Cambridge, 2015.

\bibitem{Finazzo:2014cna} 
  S.~I.~Finazzo, R.~Rougemont, H.~Marrochio and J.~Noronha,
  \emph{Hydrodynamic transport coefficients for the non-conformal quark-gluon plasma from holography}, 
  JHEP {\bf 1502}, 051 (2015). 
  
\bibitem{Nakayama:2012vs} 
  Y.~Nakayama,
 \emph{Intrinsic ambiguity in second order viscosity parameters in relativistic hydrodynamics},
  Int.\ J.\ Mod.\ Phys.\ A {\bf 27}, 1250125 (2012).

\bibitem{Abdalla:2006qj} 
  E.~Abdalla, B.~Cuadros-Melgar, A.~B.~Pavan and C.~Molina,
  \emph{Stability and thermodynamics of brane black holes}, Nucl.\ Phys.\ B {\bf 752}, 40 (2006).

\bibitem{Horowitz:1999gf} 
  G.~T.~Horowitz and N.~Itzhaki,
  \emph{Black holes, shock waves, and causality in the AdS/CFT correspondence}, 
  JHEP {\bf 9902}, 010 (1999).

\bibitem{shenker} M. Brigante, H. Liu, R.C. Myers, S. Shenker and S. Yaida,
{\it Viscosity bound violation in higher derivative gravity},
Phys.\ Rev.\ D {\bf 77}, 126006 (2008).



\bibitem{adams} J. Adams et al. [STAR Collaboration], \emph{Experimental and theoretical challenges in the
search for the quark gluon plasma: The STAR collaborationÕs critical assessment of the
evidence from RHIC collisions,} Nucl. Phys. A {\bf 757}, 102 (2005).

\bibitem{diego} D. Mateos and D. Trancanelli, \emph{Thermodynamics and Instabilities of a Strongly Coupled
Anisotropic Plasma}, \emph{JHEP} {\bf 1107}, 054 (2011).


\bibitem{Critelli:2014kra}
  R.~Critelli, S.~I.~Finazzo, M.~Zaniboni and J.~Noronha,
  \emph{Anisotropic shear viscosity of a strongly coupled non-Abelian plasma from magnetic branes}, 
  Phys.\ Rev.\ D {\bf 90},   066006 (2014).



\bibitem{rebhan} A. Rebhan and D. Steineder, \emph{Violation of the Holographic Viscosity Bound in a Strongly
Coupled Anisotropic Plasma}, Phys. Rev. Lett. {\bf 108}, 021601 (2012). 

\bibitem{Megias:2013joa} 
  E.~Megias and F.~Pena-Benitez,
  \emph{Holographic Gravitational Anomaly in First and Second Order Hydrodynamics}, 
  JHEP {\bf 1305}, 115 (2013).

\bibitem{conj} R. Emparan, H. S. Reall,  \emph{Black Holes in Higher Dimensions}, Living Rev.\ Rel.\  {\bf 11} (2008) 6.



 
 

 








  
 



\end{thebibliography}
\end{document}